# High-field Carrier Velocity and Current Saturation in Graphene Field-effect Transistors

Brett W. Scott, *Student Member, IEEE* and Jean-Pierre Leburton, *Fellow, IEEE*
Department of Electrical and Computer Engineering
and Beckman Institute for Advanced Science and Technology
University of Illinois at Urbana-Champaign
Urbana, IL 61801
e-mail: jleburto@illinois.edu

*Abstract*-We obtain the output characteristics of graphene field-effect transistors by using the charge-control model for the current, based on the solution of the Boltzmann equation in the field-dependent relaxation time approximation. Closed expressions for the conductance, transconductance and saturation voltage are derived. We found good agreement with the experimental data of Meric et al. [1], without assuming a carrier density-dependent velocity saturation.

## I. INTRODUCTION

In recent years, graphene has emerged as a novel mono-layer material with exotic physical properties for applications in high performance electronic devices [1, 2]. Namely the relation between the charge carrier energy $E$ and the two-dimensional (2D) wave vector $k = \sqrt{k_x^2 + k_y^2}$ is linear i.e. $E = \hbar v_f k$, where $v_f \sim 10^8$ cm/s is the Fermi velocity, thereby reducing the band gap to a single point (Dirac point) [3]. In this framework all carriers have a velocity with the same absolute value that is one order of magnitude larger than in conventional III-V materials [4], making graphene a promising candidate for high-speed nanoelectronics.

Recently, graphene field-effect transistors (GFETs) were successfully fabricated, and exhibited *I-V* characteristics similar to conventional silicon MOS transistors [1]. Low field mobilities were however strongly degraded by the presence of coulombic space charge in the neighboring oxides, whereas nonlinearities in the current-voltage characteristics were interpreted as caused by carrier velocity saturation for which the value would depend on the carrier concentration induced by gate voltages in the 2D graphene mono-layer.

In this paper, we developed a charge-control model for GFETs that does not require the assumption of carrier density-dependent saturation velocity to reproduce the experimental characteristics. Our model also provides closed form analytic expressions for the saturation voltage, conductance and transconductance of the device.

## II. TRANSISTOR STRUCTURE

Fig. 1 shows a schematic of the GFET, where the graphene mono-layer sits on a thick $SiO_2$ layer with capacitance $C_{back}$ on top of a back gate that controls the source and drain resistance $R_s$ at the same time as the channel threshold voltage with bias $V_{gback}$. A top gate of length $L$ separated from the graphene mono-layer by a thinner oxide with capacitance $C_{top}$ controls the carriers in the channel with $V_{gtop}$. For the sake of comparison with experiment, we will only consider p-channel device operation, but our model is valid for n-channel operation as well.

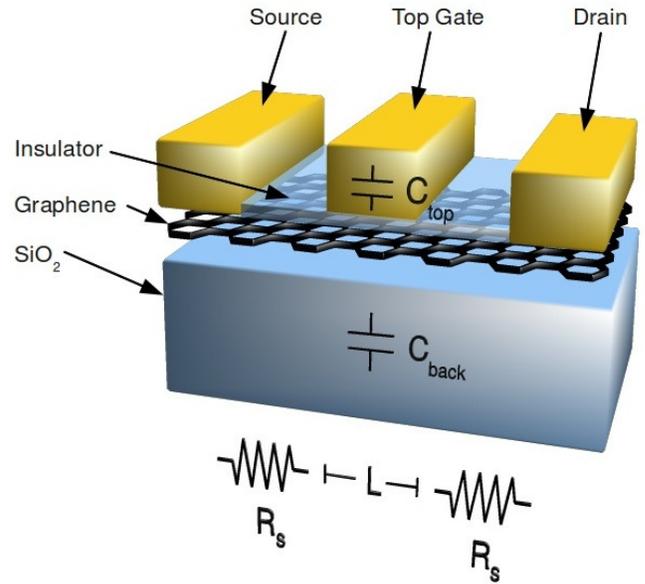

Fig. 1. Schematic of the GFET device.

## III. MODEL

In the relaxation time approximation, current in a 2D graphene layer at room temperature is diffusive and given by

$$I = \frac{2e^2}{h} W v_f \tau(p) \sqrt{\pi p} F \qquad (1)$$

where $W$ is the graphene layer width, $p$ is the hole concentration, $F$ is the electric field and $\tau(p)$ is the relaxation time for a particular carrier concentration. In the high field regime, we assume $\tau(p) = \tau_{lf}/(1 + F/F_c)$ where $F_c$ is the critical field for the onset of high energy collisions such as remote phonons, for instance, $\tau_{lf}(p) = \tau_0 (p/N_i)^{1/2}$ is the low field relaxation time dominated by scattering with charged impurities with density $N_i$, and $\tau_0$ is a time constant. By setting $\mu_0 = (e/\hbar)v_f\tau_0/\sqrt{\pi N_i}$, one recovers the conventional current expression

$$I = Wepv(F) \quad (2)$$

with $v(F) = \mu_0 F/(1 + F/F_c)$, where the low field conductance $\sigma_{lf} \propto p$, as observed experimentally.

In the charge-control model, and for $p \gg p_0$ (which is the case for all data in [1]), where $p_0$ is the minimum sheet carrier concentration, one can write $p(x) = -(C_{top}/e)[V_{g0} - V(x)]$ where $V(x)$ is the electric potential along the channel from source to drain, and $V_{g0} = V_{gtop} - V_0 < 0$. Here $V_0$ is the threshold voltage of the GFET and is defined in [1] by $V_0 = V_{gtop}^0 + (C_{back}/C_{top})(V_{gback}^0 - V_{gback})$ where $V_{gtop}^0$ and $V_{gback}^0$ are the top and back gate voltages at the Dirac point respectively. By integrating the current equation (2) from source to drain in the gradual channel approximation as in conventional MOS devices [7], and by taking into account the series resistance $R_s$ at the source and drain, we get

$$I_d = \frac{1}{4R_s}\left[V_{ds} - V_c + I_0 R_s + \sqrt{(V_{ds} - V_c + I_0 R_s) - 4I_0 R_s V_{ds}}\right] \quad (3)$$

where $V_{ds}$ is the drain-source voltage, $I_0 = (W/L)\mu_0 V_c C_{top}(V_{gtop} - V_0 - V_{ds}/2)$ and $V_c = F_c L$.

From the current expression (3), the transistor conductance $g_d$ at low drain-source bias and the saturation voltage $V_{ds(sat)}$ are readily obtained,

$$g_d(V_{ds} \to 0) = \frac{-V_{g0}}{|2R_s V_{g0} - R_c V_c|} \quad (4)$$

$$V_{ds(sat)} = \frac{2\gamma V_{g0}}{1+\gamma} + \frac{1-\gamma}{(1+\gamma)^2}\left[V_c - \sqrt{V_c^2 - 2(1+\gamma)V_c V_{g0}}\right] \quad (5)$$

while the transconductance at saturation is given by

$$g_m^{sat} = \left[1 - 1/\sqrt{1 - 2(1+\gamma)V_{g0}/V_c}\right]/(R_s + R_c) \quad (6)$$

Here $\gamma = R_s/R_c$ and $1/R_c = (W/L)\mu_0 C_{top} V_c$, so that $R_c V_c$ is independent on $V_c$, as is the conductance at low drain bias (eq. 4).

## IV. RESULTS

Eq. 4 predicts a linear relation between $1/g_{ds}$ and $1/V_{g0}$ with a slope given by $R_c V_c$ (inversely proportional to the mobility) and an asymptotic conductance value for large $V_{g0}$ reaching

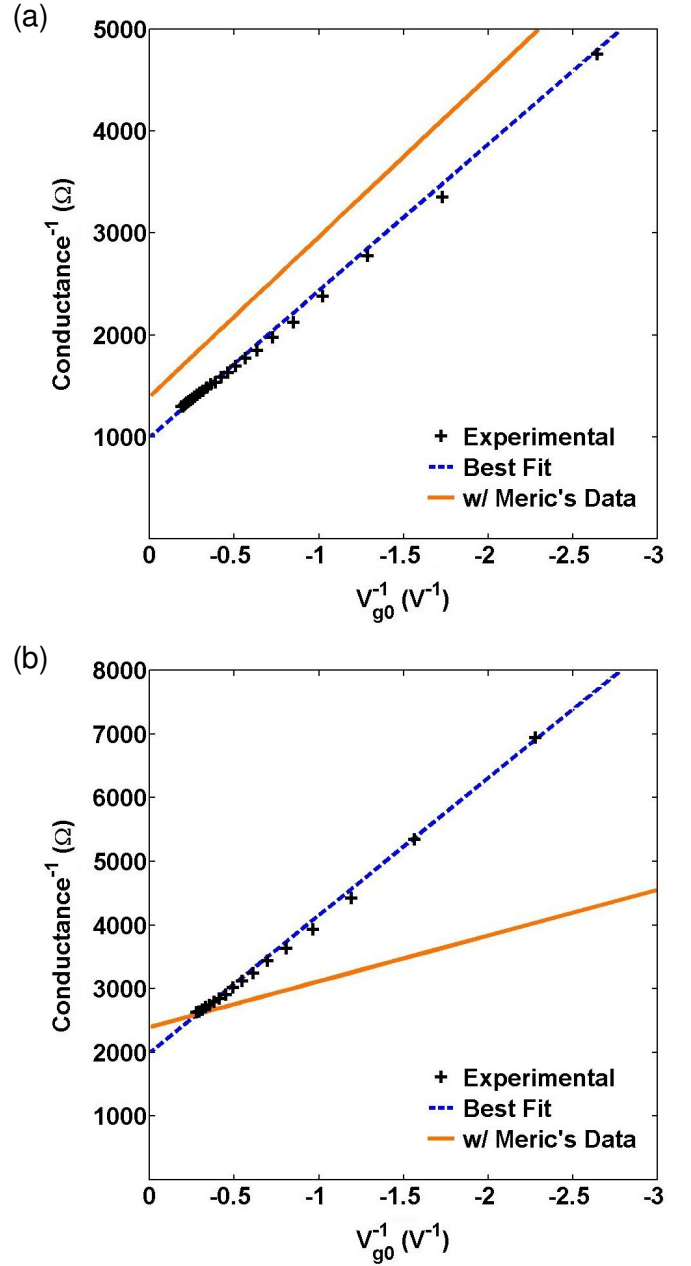

Fig. 2. Inverse of the small-signal source-drain conductance ($1/g_{ds}$) as a function of the inverse of the top gate voltage minus the threshold voltage ($1/V_{g0}$) for (a) $V_{gback} = -40$ V and (b) $V_{gback} = +40$ V.

$2R_s$. Fig. 2 shows the plots of the inverse low-bias conductance as a function of the inverse gate voltage for two substrate biases $V_{gback} = -40$ V ($V_0 = 2.36$ V; fig. 2a) and $V_{gback} = +40$ V ($V_0 = 0.64$ V; fig. 2b) in the device configuration investigated in [1]. In both cases, the experimental conductance values display the linear relation. In fig. 2a ($V_{gback} = -40$ V), the solid curve is calculated from (4) with the mobility ($\mu_0 = 550$ cm$^2$/V·s) and source resistance values ($R_s = 700$ Ω) given in [1]. One can observe the slopes of the experimental and theoretical curves are in relatively good agreement indicating similar values of mobility, while both curves are shifted from one another by a relatively constant value which is determined by the source resistance. The best fit between the two curves is obtained with $\mu_0$ = 600 cm$^2$/V·s and $R_s$ = 500 Ω, so the discrepancy is essentially due to a different value of the source resistance. In fig. 2b ($V_{gback}$ = +40 V), the disagreement between the theory ($\mu_0$ = 1200 cm$^2$/V·s; $R_s$ = 1200 Ω [1]) and experimental values is more dramatic since it affects both the slope (mobility), and to a less extent the asymptotic value of the conductance (source resistance). The best fit is obtained with $\mu_0$ = 400 cm$^2$/V·s and $R_s$ = 1000 Ω.

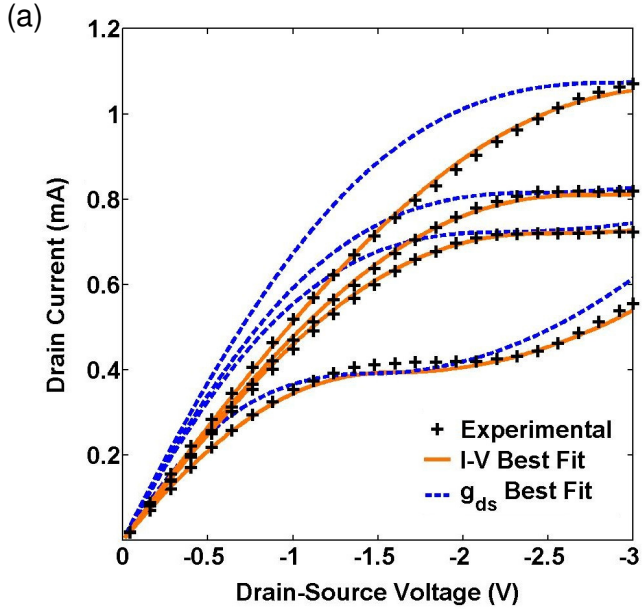

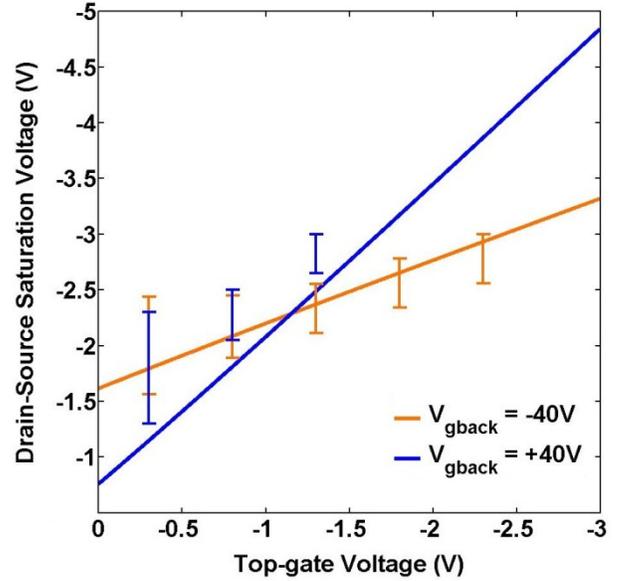

Fig. 4. Drain-source saturation voltage ($V_{ds(sat)}$) as a function of top-gate voltage ($V_{gtop}$) for two $V_{gback}$ biases.

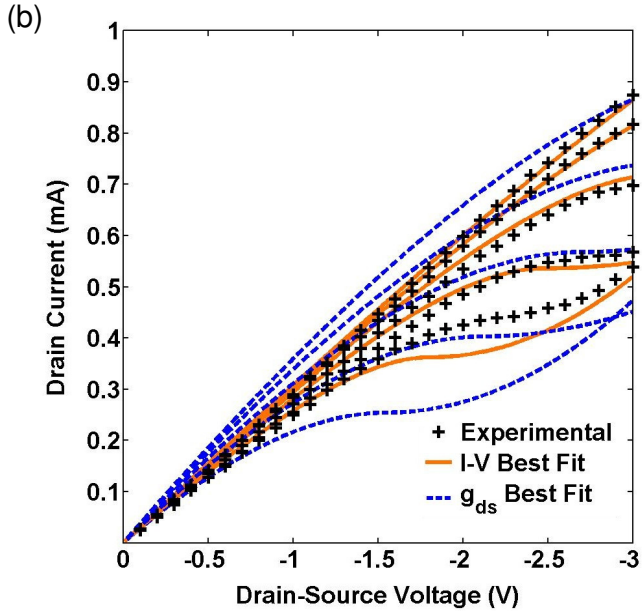

Fig. 3. Drain current ($I_d$) as a function of drain-source voltage ($V_{ds}$) for (a) $V_{gback}$ = -40V; $V_{gtop}$ = 0V, -1.5V, -1.9V and -3V (from bottom to top) and (b) $V_{gback}$ = +40V; $V_{gtop}$ = -0.8V, -1.3V, -1.8V, -2.3V and -2.8V.

In fig. 3 we display the *I-V* characteristics of the GFET for the two back gate bias conditions. In fig. 3a ($V_{gback}$ = -40 V), an excellent agreement between experiment and theory (eq. 3) is obtained with $\mu_0$ = 700 cm$^2$/V·s, $R_s$ = 800 Ω, and $V_c$ = 0.45 V for all gate biases, which provides the right current values for high (negative) $V_{ds}$. Our mobility value is 25% higher than Meric's fitted values ($\mu_0$ = 550 cm$^2$/V·s), while the source resistance is within 15% of the measured ones [1]. The up-kick in the drain current attributed to ambipolar transport for $V_{gtop}$ = 0 V is simulated by a phenomenological current term proportional to $(V_{ds}/V_{ds(sat)} - 1)^2$ [8]. For comparison, we also plot the current with the parameter values ($\mu_0$ = 600 cm$^2$/V·s and $R_s$ = 500 Ω) that fit the conductance characteristics in fig. 2a. Here we use $V_c$ = 0.5 V ($F_c$ = 5 kV/cm) for all gate biases, which gives the right current values at $V_{ds}$ = -3 V but overestimates the current at high (negative) gate and intermediate source-drain biases. Similar curves are displayed on fig. 3b ($V_{gback}$ = +40 V). Here the best fit of the *I-V* characteristics is obtained with $\mu_0$ = 1200 cm$^2$/V·s, $R_s$ = 1500 Ω and $V_c$ = 1.5 V ($F_c$ = 15 kV/cm) for all gate biases, which are also close to Meric's values [1], but significantly different from the best conductance fit on fig. 2b that underestimates (overestimates) the current at low (high) (negative) gate bias. This high value for $F_c$ compared to the GFET configuration with $V_{gback}$ = -40 V is indicative of the higher saturation voltage for similar gate biases, while the higher source resistance provides lower current than for $V_{gback}$ = -40 V, despite the higher mobility.

In fig. 4 we plot the drain-source saturation voltage as a function of gate bias (eq. 5) for the two GFET configurations. One notices the excellent agreement between theory and experiment, especially for the $V_{gback}$ = -40 V condition, whereas the discrepancy for the $V_{gback}$ = +40 V configuration is due to the uncertainty in ascertaining the experimental values that fall out of the figure.

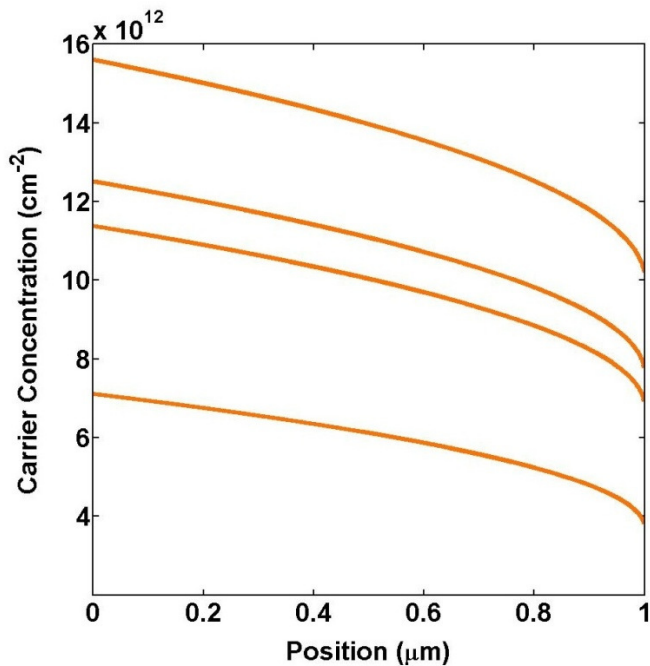

Fig. 5. Hole concentration as a function of position along the channel length (source is on the left) for $V_{gback}$ = -40 V and $V_{gtop}$ = 0 V, -1.5 V, -1.9 V and -3 V (from bottom to top) at the onset of saturation.

From the current equation (3) we obtain the expression of the potential along the channel at the onset of drain current saturation for $V_{gback}$ = -40 V and varying top-gate biases. One can observe that the channel never experiences pinchoff even for the lowest applied top-gate bias (0 V) since according to [1], the minimum sheet carrier density $n_0 \sim 5 \times 10^{11}$ cm$^{-2}$, which also validates the assumption of unipolar transport.

## V. DISCUSSION AND CONCLUSIONS

We provide a coherent model that computes the output and transfer characteristics of GFETs for two back-gate bias conditions, for which the source and drain contacts are either p- or n-types. For unipolar transport, closed form expressions are obtained for the current, low drain bias conductance, transconductance and saturation voltages, which rely on three parameters i.e. low field carrier mobility, source-drain resistance and critical field for the high energy carrier scattering, to reproduce the experimental I-V characteristics for each back gate condition. In particular we predict a linear dependence of the inverse low-field conductance versus inverse gate voltage, which is qualitatively verified, and we point out a discrepancy between the parameter values used for the $g_{ds}$ -$V_{g0}$ plots and the I-V plots, especially for positive back gate voltage, which has not been resolved so far. However the predicted quasi-linear dependence between saturation voltage and gate voltage is well reproduced experimentally.

Let us emphasize that our model relies on only one $F_c$ parameter to describe the current at high drain biases for all top gate biases, which according to the velocity field relation $v(F)$ implies a single saturation velocity $v_{sat}$ = 3.2 x 10$^6$ cm/s (1.8 x 10$^7$ cm/s) for $V_{gback}$ = -40 V (+40 V), unlike Meric's model that requires a concentration dependent saturation velocity to fit the experimental data. In this respect, let us point out that close analysis of the source–drain field profile indicates that the maximum fields achieved in the highest drain biases are only a few times the critical field values $F_c$, which is far from achieving saturation; it is therefore quite possible that the velocity-field relation acquires a lower slope due to remote phonon scattering rather than saturating.

Finally, we also point out that detailed analysis of the charge control model indicates that even for the lowest (negative) top gate bias i.e. $V_{gtop}$ = 0 V (-0.8 V) for $V_{gback}$ = -40 V (+40 V), the channel never pinches–off, which suggests that the current increase at high drain biases may be due to other causes than electron injection.

ACKNOWLEDGMENT

B.S. would like to thank the Department of Electrical and Computer Engineering at the University of Illinois at Urbana-Champaign for their financial support.

REFERENCES

[1] I. Meric et al., "Current saturation in zero-bandgap, top-gated graphene field-effect transistors" *Nature Nanotechnology*, vol. 3, pp. 654-659, November 2008.
[2] F. Xia, D.B. Farmer, Y-M. Lin, and P. Avouris, "Graphene Field-Effect Transistors with High On/Off Current Ratio and Large Transport Band Gap at Room Temperature," *Nano Lett.*, vol. 10, pp. 715-718, January 2010.
[3] M.I. Katsnelson, "Graphene: carbon in two dimensions," *Materials Today*, vol. 10, pp. 20-27, 2007.
[4] E. Kartheuser, *Polarons in Ionic Crystals and Polar Semiconductors*, NATO Advanced Study Institute, Antwerp, pp. 717-733, 1971.
[5] K. Hess, "Advanced Theory of Semiconductor Devices" Wiley-IEEE Press, New York, 2000.
[6] J.H. Chen et al., "Diffusive charge transport in graphene on SiO$_2$," *Solid State Communications*, vol. 149, pp. 1080-1086, July 2009.
[7] R.S. Muller and T.I. Kamins, "Device Electronics for Integrated Circuits" 3$^{rd}$ edition, John Wiley and Sons, New York, 2003.
[8] A. Girdhar and J.P. Leburton, unpublished.